\documentclass[twocolumn]{aastex62}

\newcommand{\kms}{km~s$^{-1}$}
\newcommand{\msun}{$M_{\odot}$}

\newcommand{\mas}{mas~yr$^{-1}$}

\shortauthors{Brown et al.}
\accepted{Aug 17, 2018}
\begin{document}

\title{Gaia and the Galactic Center Origin of Hypervelocity Stars}

\correspondingauthor{Warren R.\ Brown}
\email{wbrown@cfa.harvard.edu}

\author[0000-0002-4462-2341]{Warren R.\ Brown}
\affil{Smithsonian Astrophysical Observatory\\
60 Garden St, Cambridge, MA 02138, USA}

\author{Mario G.\ Lattanzi} \affil{Instituto Nazionale di Astrofisica -- 
Osservatorio Astronomico di Torino\\
Via Osservatorio 20, I-10025 Pino Torinese, Italy}

\author{Scott J.\ Kenyon}
\affil{Smithsonian Astrophysical Observatory\\
60 Garden St, Cambridge, MA 02138, USA}

\author{Margaret J.\ Geller}
\affil{Smithsonian Astrophysical Observatory\\
60 Garden St, Cambridge, MA 02138, USA}

\begin{abstract}

	We use new {\it Gaia} measurements to explore the origin of the highest 
velocity stars in the Hypervelocity Star Survey.  The measurements reveal a clear 
pattern in the B-type stars.  Halo stars dominate the sample at speeds $\simeq$100 
\kms\ below Galactic escape velocity.  Disk runaway stars have speeds up to 
$\simeq$100 \kms\ above Galactic escape velocity, but most disk runaways are bound.  
Stars with speeds $\gtrsim$100 \kms\ above Galactic escape velocity originate from 
the Galactic center.  Two bound stars may also originate from the Galactic center.  
Future {\it Gaia} measurements will enable a large, clean sample of Galactic center 
ejections for measuring the massive black hole ejection rate of hypervelocity stars, 
and for constraining the mass distribution of the Milky Way dark matter halo.

\end{abstract}


\section{Introduction}

	\citet{hills88} first proposed that a 3-body exchange between a pair of
stars and a massive black hole (MBH) can eject ``hypervelocity stars'' (HVSs) at
1000 \kms\ velocities from the Galactic center.  We discovered the first HVS
\citep{brown05}.  This 3~\msun\ main sequence B star moves with a Galactic rest
frame velocity $>$670 \kms, about twice Galactic escape velocity at its current
distance of 100 kpc.  Only a gravitational interaction with a massive compact object
can plausibly explain its motion.

	The discovery of HVS1 inspired the HVS Survey, a targeted radial velocity
survey of B-type stars that should not exist at faint magnitudes in the halo
\citep{brown06, brown07b}.  Twenty one stars are significantly unbound in radial
velocity alone \citep{brown14}.  The extreme velocities, the short-lived nature of
the stars, their distribution in Galactic latitude, and their overall numbers match
theoretical expectations for the Galactic center origin proposed by \citet{hills88}.  
However the measurements provide only an indirect link to the MBH.

	Alternative origins for unbound stars include ejection from the Galactic 
disk through binary disruption \citep{blaauw61, poveda67} and ejection from the 
Large Magellanic Cloud \citep{boubert16, boubert17}.  Galactic disk ejections are 
called ``runaways'' \citep{blaauw61, greenstein74}.  The disruption of a binary by a 
supernova, where the surviving star is released at the orbital velocity of the 
progenitor binary, can yield unbound runaways in extreme circumstances 
\citep[e.g.][]{tauris15}.  The first example of an unbound main sequence
runaway is the B star HD~271791 \citep{heber08, przybilla08c}.  The first example of 
an unbound Large Magellanic Cloud ejection is the B star HE~0437$-$5439 
\citep{edelmann05, przybilla08, erkal18}.  Compact objects like white dwarfs can 
have higher binary disruption ejection velocities than main sequence stars.  The 
first observational examples are the unbound subdwarf O star US~708 \citep{hirsch05, 
justham09, wang09, geier15}, the white dwarf LP~$40-365$ \citep{vennes17, raddi18}, 
and three white dwarf candidates found with {\it Gaia} \citep{shen18}.

	The European Space Agency mission {\it Gaia} has begun a new era of
precision astrometry.  The trajectories of unbound stars hold the key to their
origin.  Measuring radial velocity to \kms\ precision is easy with modern
spectroscopy; measuring tangential velocity, the product of distance and proper
motion, is difficult.  Known hypervelocity stars are at distances of 50 to 100 kpc;
their expected proper motions are $<$1 \mas.  Newly released {\it Gaia} Data Release
2 provides improved proper motions for many HVSs \citep{gaia18}.  Here, we use {\it
Gaia} measurements to determine the origin of stars from the HVS Survey
\citep{brown07b, brown14} on the basis of their trajectory and velocity.

	In Section 2 we define the sample and compare {\it Gaia} proper motions with
previous {\it Hubble Space Telescope} {\it (HST)} measurements.  In Section 3 we
evaluate the origin of these stars on the basis of computed trajectories and
ejection velocities.  The results are in Section 4, and we conclude in Section 5.  
We identify Galactic center HVSs, Galactic disk runaways, and Galactic halo stars
with different but overlapping velocities; the highest velocity stars are probably
Galactic center ejections.

% TABLE 1
\begin{deluxetable*}{lcccccc}
\tabletypesize{\scriptsize}
\tablecaption{HVS Survey Stars with $v_{rf}>+275$ \kms, Ordered by $v_{rf}$\label{tab:sample}}
\tablewidth{0pt}
\tablecolumns{7}
\tablehead{
  \colhead{ID} & \colhead{RA} & \colhead{Dec} & \colhead{$g$} &
  \colhead{$Gaia~(\mu_\alpha, \mu_\delta)$} &
  \colhead{$HST~(\mu_\alpha, \mu_\delta)$} &
  \colhead{GPS1 $(\mu_\alpha, \mu_\delta)$} \\
  \colhead{} & \colhead{(J2000)} & \colhead{(J2000)} & \colhead{(mag)} &
  \colhead{(\mas, \mas )} & \colhead{(\mas, \mas )} & \colhead{(\mas, \mas )} 
}
        \startdata
HVS1	&  9:07:45.0 &  2:45:07  & 19.79 & $-1.012\pm1.321$, $-0.269\pm0.879$ & $ 0.080\pm0.261$, $-0.117\pm0.221$ &                 \nodata            \\
HVS5	&  9:17:59.5 & 67:22:38  & 17.93 & $ 0.017\pm0.176$, $-1.164\pm0.268$ & $ 0.554\pm0.615$, $-0.438\pm0.589$ & $-1.265\pm2.248$, $ 3.747\pm1.837$ \\
HVS4	&  9:13:01.0 & 30:51:20  & 18.40 & $-0.308\pm0.647$, $-1.055\pm0.481$ & $-0.230\pm0.362$, $-0.422\pm0.358$ & $-1.378\pm2.128$, $ 2.240\pm1.651$ \\
HVS6	& 11:05:57.5 &  9:34:39  & 19.06 & $-0.367\pm0.664$, $-0.694\pm0.507$ & $ 0.051\pm0.568$, $ 0.307\pm0.967$ & $ 2.791\pm1.773$, $ 2.488\pm2.369$ \\
HVS19	& 11:35:17.8 &  8:02:01  & 20.18 & $-0.626\pm1.790$, $ 0.363\pm1.224$ &                 \nodata            &                 \nodata            \\
HVS22	& 11:41:46.4 &  4:42:17  & 20.26 & $ 0.180\pm2.024$, $ 1.964\pm1.443$ &                 \nodata            &                 \nodata            \\
HVS9	& 10:21:37.1 & -0:52:35  & 18.84 & $ 0.345\pm0.743$, $-0.117\pm0.747$ & $-1.260\pm0.736$, $-0.250\pm0.697$ & $ 0.212\pm1.427$, $ 0.439\pm1.393$ \\
HVS18	& 23:29:04.9 & 33:00:11  & 19.66 & $-0.308\pm0.656$, $-0.157\pm0.495$ &                 \nodata            & $-4.434\pm2.749$, $ 5.957\pm3.858$ \\
B733	& 14:49:55.6 & 31:03:51  & 15.75 & $-1.231\pm0.060$, $-4.547\pm0.094$ & $-1.769\pm0.939$, $-3.709\pm1.017$ & $ 1.425\pm1.276$, $-1.627\pm1.102$ \\
HVS17	& 16:41:56.4 & 47:23:46  & 17.50 & $-1.069\pm0.198$, $-1.104\pm0.323$ &                 \nodata            & $ 0.615\pm1.763$, $ 0.314\pm1.551$ \\
HVS13	& 10:52:48.3 & -0:01:34  & 20.18 & $-0.729\pm1.949$, $ 0.047\pm1.345$ & $-0.898\pm0.385$, $ 0.456\pm0.439$ &                 \nodata            \\
HVS12	& 10:50:09.6 &  3:15:51  & 19.77 & $ 0.425\pm1.377$, $ 0.193\pm0.993$ & $-0.404\pm0.364$, $ 0.314\pm0.337$ & $-3.040\pm2.368$, $-0.678\pm2.408$ \\
HVS10	& 12:03:37.9 & 18:02:50  & 19.30 & $-3.161\pm1.288$, $-1.149\pm0.494$ & $-1.075\pm0.362$, $-0.583\pm0.419$ & $ 3.292\pm2.459$, $ 0.702\pm1.933$ \\
HVS8	&  9:42:14.0 & 20:03:22  & 18.05 & $-0.805\pm0.365$, $-0.055\pm0.369$ & $-0.821\pm1.261$, $-0.039\pm0.697$ & $ 3.217\pm2.425$, $-0.251\pm2.582$ \\
HVS7	& 11:33:12.1 &  1:08:25  & 17.75 & $-0.701\pm0.373$, $ 0.412\pm0.253$ & $ 0.996\pm0.961$, $-0.549\pm1.158$ & $-4.776\pm1.377$, $ 0.717\pm1.440$ \\
HVS20	& 11:36:37.1 &  3:31:07  & 19.89 & $ 0.458\pm1.451$, $ 0.574\pm1.014$ &                 \nodata            &                 \nodata            \\
HVS21	& 10:34:18.3 & 48:11:35  & 19.78 & $ 0.003\pm0.693$, $-0.224\pm0.881$ &                 \nodata            &                 \nodata            \\
B485	& 10:10:18.8 & 30:20:28  & 16.16 & $-0.789\pm0.131$, $-0.141\pm0.127$ & $-1.665\pm0.722$, $-1.149\pm0.628$ & $-0.820\pm2.048$, $-0.802\pm1.977$ \\
HVS24	& 11:11:36.4 &  0:58:56  & 18.98 & $ 0.292\pm0.777$, $-0.379\pm0.578$ &                 \nodata            & $-4.431\pm2.916$, $-1.653\pm3.129$ \\
HVS16	& 12:25:23.4 &  5:22:34  & 19.40 & $-1.643\pm1.518$, $-1.101\pm0.856$ &                 \nodata            & $ 1.742\pm2.090$, $-2.240\pm1.903$ \\
B1080	& 10:33:57.3 & -1:15:07  & 18.73 & $-0.957\pm0.599$, $-0.619\pm0.417$ &                 \nodata            &                 \nodata            \\
HVS15	& 11:33:41.1 & -1:21:14  & 19.24 & $-0.888\pm1.291$, $-0.316\pm0.567$ &                 \nodata            & $-2.524\pm1.783$, $-4.246\pm1.778$ \\
B1085	& 11:22:55.8 & -9:47:35  & 17.53 & $-2.251\pm0.246$, $-0.333\pm0.172$ &                 \nodata            & $ 2.116\pm2.201$, $-3.046\pm1.861$ \\
B434	& 11:02:24.4 &  2:50:03  & 18.15 & $ 0.095\pm0.375$, $-1.954\pm0.300$ & $-1.613\pm0.575$, $-0.264\pm0.650$ & $ 1.006\pm1.656$, $-1.559\pm1.644$ \\
B537	&  0:28:10.3 & 21:58:10  & 17.35 & $ 0.761\pm0.229$, $-0.506\pm0.120$ &                 \nodata            &                 \nodata            \\
B080	& 11:06:28.2 & -8:22:48  & 18.68 & $-0.186\pm0.537$, $ 0.060\pm0.415$ &                 \nodata            & $-2.575\pm3.455$, $-4.762\pm2.080$ \\
B572	&  0:59:56.1 & 31:34:39  & 18.02 & $ 0.488\pm0.329$, $-0.989\pm0.366$ &                 \nodata            & $-1.850\pm4.124$, $ 2.423\pm1.393$ \\
B458	& 10:43:18.3 & -1:35:03  & 19.44 & $ 0.652\pm0.892$, $-0.197\pm0.747$ &                 \nodata            & $-0.276\pm1.559$, $-2.714\pm1.648$ \\
B711	& 14:20:01.9 & 12:44:05  & 17.00 & $ 0.594\pm0.209$, $-2.582\pm0.172$ & $-0.960\pm0.942$, $ 1.545\pm0.999$ & $ 2.790\pm1.527$, $ 0.636\pm1.140$ \\
B576	& 14:04:32.4 & 35:22:58  & 17.53 & $-3.201\pm0.130$, $-0.957\pm0.129$ &                 \nodata            & $ 3.199\pm1.354$, $ 0.729\pm1.379$ \\
B095	& 10:13:59.8 & 56:31:12  & 19.86 & $-0.319\pm0.854$, $ 0.769\pm0.864$ &                 \nodata            & $ 1.972\pm2.280$, $ 0.282\pm1.693$ \\
B495	& 11:52:45.9 & -2:11:16  & 18.22 & $-0.124\pm0.505$, $ 0.164\pm0.227$ &                 \nodata            & $-0.604\pm1.648$, $ 0.621\pm1.325$ \\
B1139	& 18:00:50.9 & 48:24:25  & 17.66 & $-1.351\pm0.192$, $-1.032\pm0.214$ &                 \nodata            & $-0.843\pm1.132$, $ 0.216\pm1.235$ \\
B598	& 14:17:23.3 & 10:12:46  & 18.49 & $-1.925\pm0.583$, $-0.820\pm0.526$ &                 \nodata            & $-1.156\pm2.400$, $-2.482\pm2.407$ \\
B329	& 15:48:06.9 &  9:34:24  & 19.05 & $-1.321\pm0.603$, $-0.983\pm0.583$ &                 \nodata            & $-0.367\pm2.016$, $-3.864\pm1.859$ \\
B129	&  7:49:50.2 & 24:38:41  & 18.63 & $ 0.697\pm0.583$, $-0.680\pm0.432$ &                 \nodata            & $ 0.548\pm1.391$, $-1.619\pm1.486$ \\
B143	&  8:18:28.1 & 57:09:22  & 17.55 & $ 0.381\pm0.207$, $-1.490\pm0.180$ &                 \nodata            & $ 4.820\pm1.740$, $ 2.803\pm1.555$ \\
B481	& 23:22:29.5 &  4:36:51  & 17.63 & $ 2.027\pm0.322$, $-1.321\pm0.209$ &                 \nodata            & $-1.861\pm1.506$, $ 0.834\pm1.475$ \\
B167	&  9:07:10.1 & 36:59:58  & 18.16 & $-0.630\pm0.310$, $-0.178\pm0.293$ &                 \nodata            & $ 2.015\pm1.809$, $-0.537\pm1.497$ \\
	\enddata
\end{deluxetable*}

\section{Data}

\subsection{Sample}

	We study 42 radial velocity outliers from the Hypervelocity Star Survey 
\citep{brown07b, brown14}.  We include all stars with heliocentric radial velocity 
transformed to the Galactic frame $v_{rf}>$+275 \kms, 
	\begin{equation} v_{rf} = v_{helio} + 11.1\cos{l}\cos{b} +
247.24\sin{l}\cos{b} + 7.25\sin{b},\end{equation} where $l$ and $b$ are Galactic
longitude and latitude, respectively, and we assume the Sun is moving with respect
to the local standard of rest as measured by \citet{schonrich10} and the Galactic
disk circular velocity is 235 \kms\ \citep{reid14}.  We choose $v_{rf}>$+275 \kms\
because of the significant absence of negative velocity outliers in the HVS Survey.  
The four most negative velocity stars have $-300<v_{rf}<-275$ \kms\ \citep{brown14},
implying that the 42 stars with $v_{rf}>+275$ \kms\ are a relatively clean sample of
ejected stars with minimal halo star contamination.

	The HVS Survey targeted stars selected by color \citep{brown12b} with no
kinematic selection.  The spectroscopy is 99\% complete.  Stellar atmosphere fits
establish that the majority of unbound stars are main sequence B stars \citep[and
references therein]{brown14, brown15a}.  The bound stars are also probably main
sequence B stars on the basis of their velocity distribution.  The absence of $-300$
\kms\ stars in the HVS Survey implies that the $+300$ \kms\ stars must have
lifetimes less than their $\sim$1 Gyr orbital turn-around time \citep{brown07a,
kollmeier07, yu07}.  Thus they are B stars, and we calculate stellar distances using
Padova main-sequence tracks \citep{girardi04, marigo08, bressan12} with an estimated
precision of 15\%.  We transform heliocentric distances to the Galactic frame
assuming that the Sun is located 8 kpc from the Galactic center \citep{camarillo18}.

	The HVS Survey exclusively samples the stellar halo.  The $17<g<20.25$
apparent magnitude limit corresponds to approximately $30<R<120$ kpc in
Galactocentric radial distance.  For completeness, we include five stars from the
bright $15<g<19.5$ portion of the initial HVS Survey \citep{brown07b}.  The bright
stars are nearby $10<R<30$ kpc and bound, but four have {\it HST} proper motion
measurements for comparison with {\it Gaia}.

	Table \ref{tab:sample} lists the sample of 42 stars with their coordinates
and apparent $g$-band SDSS magnitude.  We sort the table by decreasing $v_{rf}$.  
We refer to bound objects with the letter B followed by their target number in the
HVS Survey, and unbound objects by their published HVS number.  We also list the
east-west and north-south components of proper motion, ($\mu_{\alpha}$,
$\mu_{\delta}$), obtained from three sources.

\subsection{Proper Motions}

	{\it Gaia} Data Release 2 contains proper motions for 39 of the 42 stars 
listed in Table \ref{tab:sample}.  The three missing stars (HVS14, HVS23, and B149)  
have too few {\it Gaia} measurements for a robust solution.  We drop them from 
further consideration.  The remaining 39 velocity outliers satisfy the quality 
controls recommended by \citet{lindegren18}: i.e.\ the objects all have $>10$ 
visibility periods, $<1.4$ mas astrometric excess noise, and the longest semi-major 
axis in the 5-dimensional error ellipses is $<1.9$ mas.  For reference, the median 
{\it Gaia} proper motion error for the 39 objects is $\pm0.73$ \mas.  The brightest 
stars have errors of only $\pm$0.11 \mas.

\subsubsection{{\it Gaia} Covariances}

	Because {\it Gaia} values are derived from a 5-parameter astrometric 
solution \citep{lindegren18}, the {\it Gaia} proper motions are correlated with our 
choice of parallax = 1/distance.  We simplify the issue by assuming that position 
errors are zero.  The remaining three covariances are between parallax and the 
two components of proper motion.

	In practice, {\it Gaia} parallax errors provide so little constraint that 
the covariances affect our results very little.  {\it Gaia} parallaxes for our 
sample range from -1.06 mas to +1.20 mas, with a mean of $-0.011 \pm 0.43$ mas.  
The mean spectrophotometric parallax in our sample, $0.02\pm0.003$ mas, is 100 times 
more precise.  The typical 0.5-$\sigma$ difference between the {\it Gaia} parallax 
and our spectrophotometric parallax yields a typical 0.055 \mas\ shift in the {\it 
Gaia} proper motion value, a shift that is about 10\% of the published proper motion 
error.  Physically, this shift corresponds to a 13 \kms\ ($\simeq$10\%) change in 
tangential velocity, and an even smaller change to the total velocity of the stars.

	Tables \ref{tab:sample} lists the {\it Gaia} proper motion values 
appropriate for our adopted distances.

\subsubsection{HST Comparison}

	\citet{brown15a} publish independent proper motion measurements made with
{\it HST} for 14 stars.  The {\it HST} proper motions were measured relative to
background galaxies.  They thus provide a test of {\it Gaia}'s absolute reference
frame.

\begin{figure}          % FIGURE:  PM COMPARISON
    \includegraphics[width=3.5in]{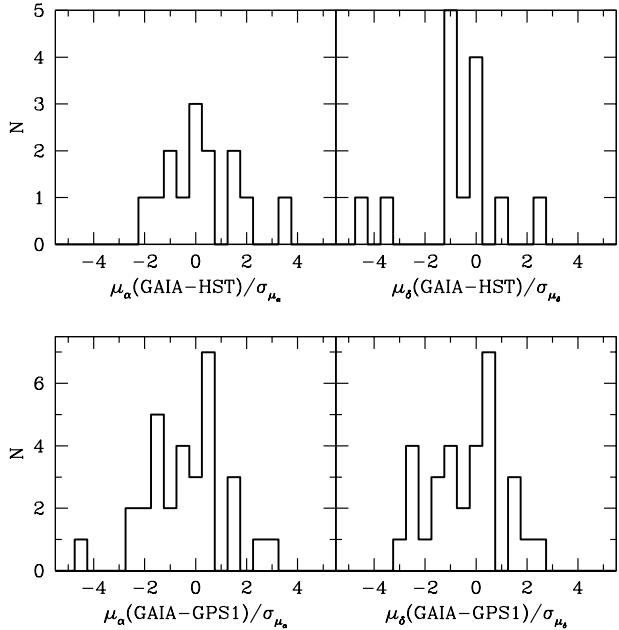}
    \caption{ \label{fig:comparison}
	Difference between {\it Gaia} and {\it HST} proper motions (upper panel) or 
GPS1 proper motions (lower panel), divided by the errors summed in quadrature.  
Left-hand panels plot east-west ($\mu_\alpha$) differences; right-hand panels plot 
north-south ($\mu_\delta$) differences.  The mean differences are consistent with 
zero, but the comparison finds notable outliers.
	}
 \end{figure}

	Figure \ref{fig:comparison} (upper panel) plots the difference between {\it
Gaia} and {\it HST} proper motion divided by the errors summed in quadrature.  
Clipping two $>$3$\sigma$ outliers (see below), the average difference $\Delta
\mu_{\alpha}(Gaia - HST) = -0.16 \pm0.31$ \mas\ and $\Delta \mu_{\delta}(Gaia - HST)
= -0.20 \pm 0.25$ \mas\ consistent with zero.

	The comparison also reveals that the most problematic {\it HST} measurements
are the brightest stars.  For bright stars, \citet{brown15a} paired short and long
exposures to tie the stars to faint background galaxies.  This approach likely
introduced additional error.  We add $\pm$0.5 \mas\ in quadrature to the published
{\it HST} error for those objects (B434, B485, B711, B711, HVS7, HVS8).  The median
{\it HST} proper motion error for the 14 stars is then $\pm0.91$ \mas; the
best-measured star has an error of $\pm0.34$ \mas.

	The {\it HST} measurements highlight the value of obtaining pointed
observations with long exposure times.  Although {\it Gaia} errors are 3 times
better than {\it HST} errors for bright stars, {\it HST} errors are 4 times better
than {\it Gaia} errors for faint stars like HVS1.  Errors are comparable in size at
$g\simeq18.5$ mag.  A weighted mean would maximize the information from {\it HST}
and {\it Gaia}, however we do not want to add measurements that include possible
systematic error.

	We adopt {\it Gaia} proper motions for the seven bright $g<18$ stars with
$\sigma_{HST} > 2\sigma_{Gaia}$.  This subset includes all stars observed with
paired short+long exposures in the {\it HST} program.  We adopt a weighted mean for
the three $18<g<19$ stars where $\sigma_{HST}$ and $\sigma_{Gaia}$ are within a
factor of two (HVS4, HVS6, and HVS9).  We adopt the {\it HST} proper motions for the
four $g>19$ stars with $\sigma_{Gaia} > 2\sigma_{HST}$ (HVS1, HVS10, HVS12, HVS13).

\subsubsection{GPS1 Comparison}

	We compare {\it Gaia} proper motions with the Gaia-PanStarrs1-SDSS catalog
\citep[GPS1,][]{tian17}.  GPS1 proper motions are based on astrometric positions
from ground-based PanStarrs1, SDSS, and 2MASS catalogs plus {\it Gaia} Data Release
1.  Because the GPS1 time baseline comes from the ground-based catalogs, and because
{\it Gaia} Data Release 2 is a new solution to the {\it Gaia} measurements, GPS1
proper motions are essentially independent of {\it Gaia} Data Release 2 values.  
GPS1 proper motions are available for 33 stars.

	Figure \ref{fig:comparison} (lower panel) plots the difference between {\it 
Gaia} and {\it GPS1} proper motion divided by the errors summed in quadrature.  
Clipping two $>$3$\sigma$ outliers, the average difference $\Delta \mu_{RA}(Gaia - 
{\rm GPS1}) = -0.18 \pm0.40$ \mas\ and $\Delta \mu_{Dec}(Gaia - {\rm GPS1}) = -0.38 
\pm 0.35$ \mas\ consistent with zero.  The distribution in declination shows a 
possible asymmetry.

	Despite the longer time baseline of ground-based observations, $\sigma_{\rm
GPS1} \simeq 5 \sigma_{Gaia}$.  Thus we do not use GPS1 values here.  The median
GPS1 proper motion error for the 33 stars is $\pm2.69$ \mas; the best-measured star
has an error of $\pm1.68$ \mas.

\section{Analysis}

	We evaluate the probability of origin on the basis of computed trajectories
and ejection velocities.  We select a gravitational potential model, trace the
trajectory of each star backwards in time, and calculate the required ejection
velocity from that position in the Milky Way.  We estimate likelihood from the
density distribution of simulated trajectories as they cross the Galactic plane.

\subsection{Gravitational Potential Model}

	To address the origin of HVS ejections from the Galactic center, we require
a gravitational potential model that fits observed mass measurements from the
Galactic center to the outer halo.  We adopt the three component bugle-disk-halo
model of \citet{kenyon14}.  The model has a flat 235 \kms\ rotation curve and a
$10^{12}$ \msun\ halo mass consistent with recent {\it Gaia} measurements from the
orbits of globular clusters and dwarf galaxies \citep{gaia18b, fritz18, posti18,
watkins18}.

	The results are insensitive to the choice of potential model because the 
stars are on nearly radial trajectories.  Inserting a simulated 
$10^{11}$~\msun\ Large Magellanic Cloud into the potential model 
\citep[see][]{kenyon18} changes the computed flight times by $<$1 Myr, changes the 
Galactic plane crossing location of the orbits by $<$0.4 kpc, and changes the 
effective escape velocity by $<$10 \kms.  We thus choose to work with the 
3-component model.

	We determine effective Galactic escape velocity, $v_{esc}$, by dropping a 
test particle from rest at the virial radius.  At the Solar circle $R=8$ kpc, 
$v_{esc}=580$ \kms\ consistent with the most recent Solar neighborhood escape 
velocity measurement \citep{monari18}.  At the radius of influence of the 
supermassive black hole, $v_{esc}\ge900$ \kms\ \citep{kenyon08}.  Only a 
gravitational interaction with the supermassive black hole can eject a main sequence 
star at $>$900 \kms\ \citep{hills88}.  At the median $R=55$ kpc depth of the HVS 
Survey sample, $v_{esc}=350$ \kms.

 \begin{figure*}          % FIGURE:  Proper Motions
%    \noindent \includegraphics[width=3.5in]{/pool/wbrown0/GAIA/Prelim/pm1.ps}
%	       \includegraphics[width=3.5in]{/pool/wbrown0/GAIA/Prelim/pm2.ps} \\
%    \noindent \includegraphics[width=3.5in]{/pool/wbrown0/GAIA/Prelim/pm3.ps}
%	       \includegraphics[width=3.5in]{/pool/wbrown0/GAIA/Prelim/pm4.ps} \\
%    \centerline{ \includegraphics[width=3.5in]{/pool/wbrown0/GAIA/Prelim/pm5.ps} }
    \noindent \includegraphics[width=3.5in]{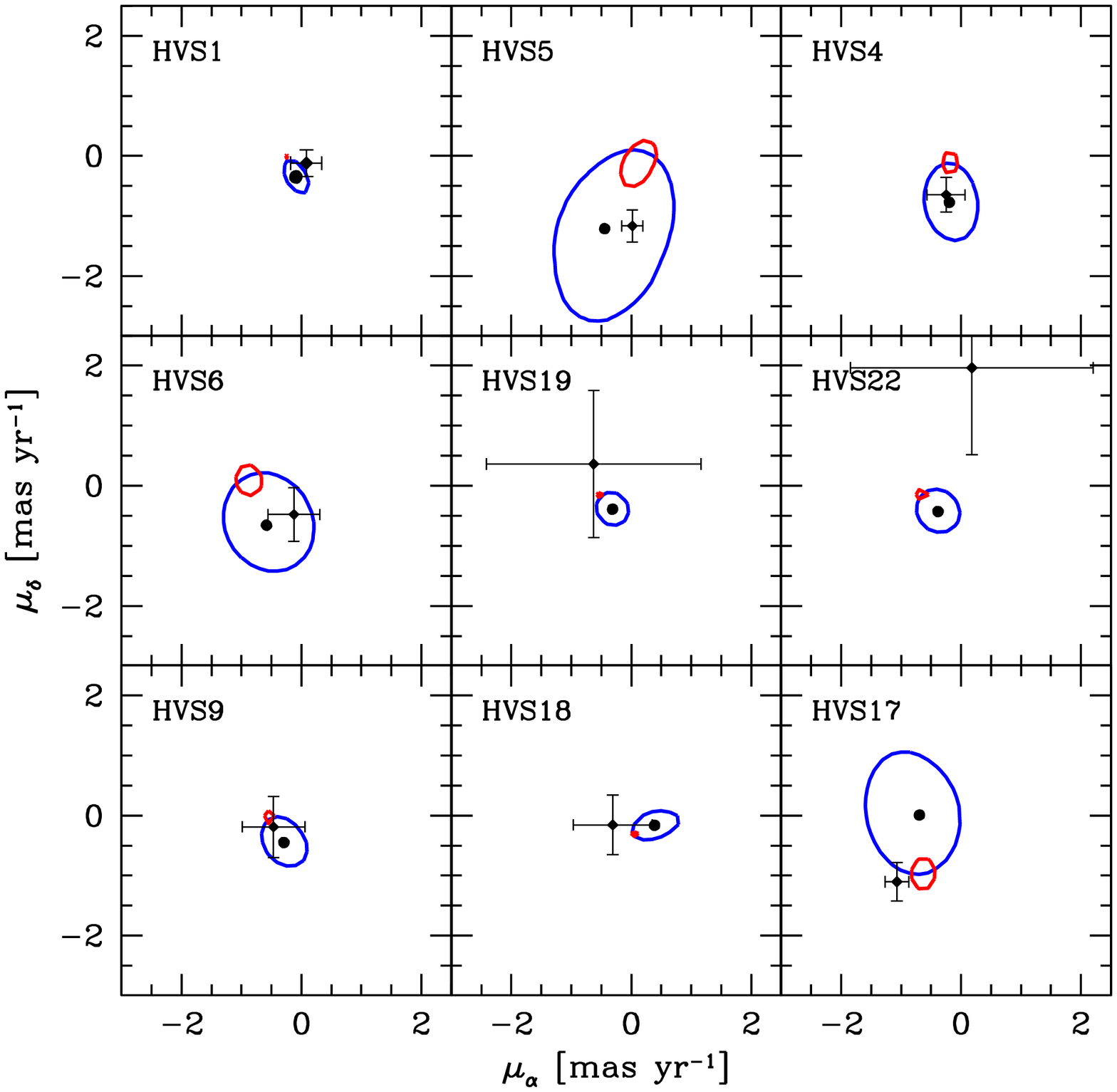}
	      \includegraphics[width=3.5in]{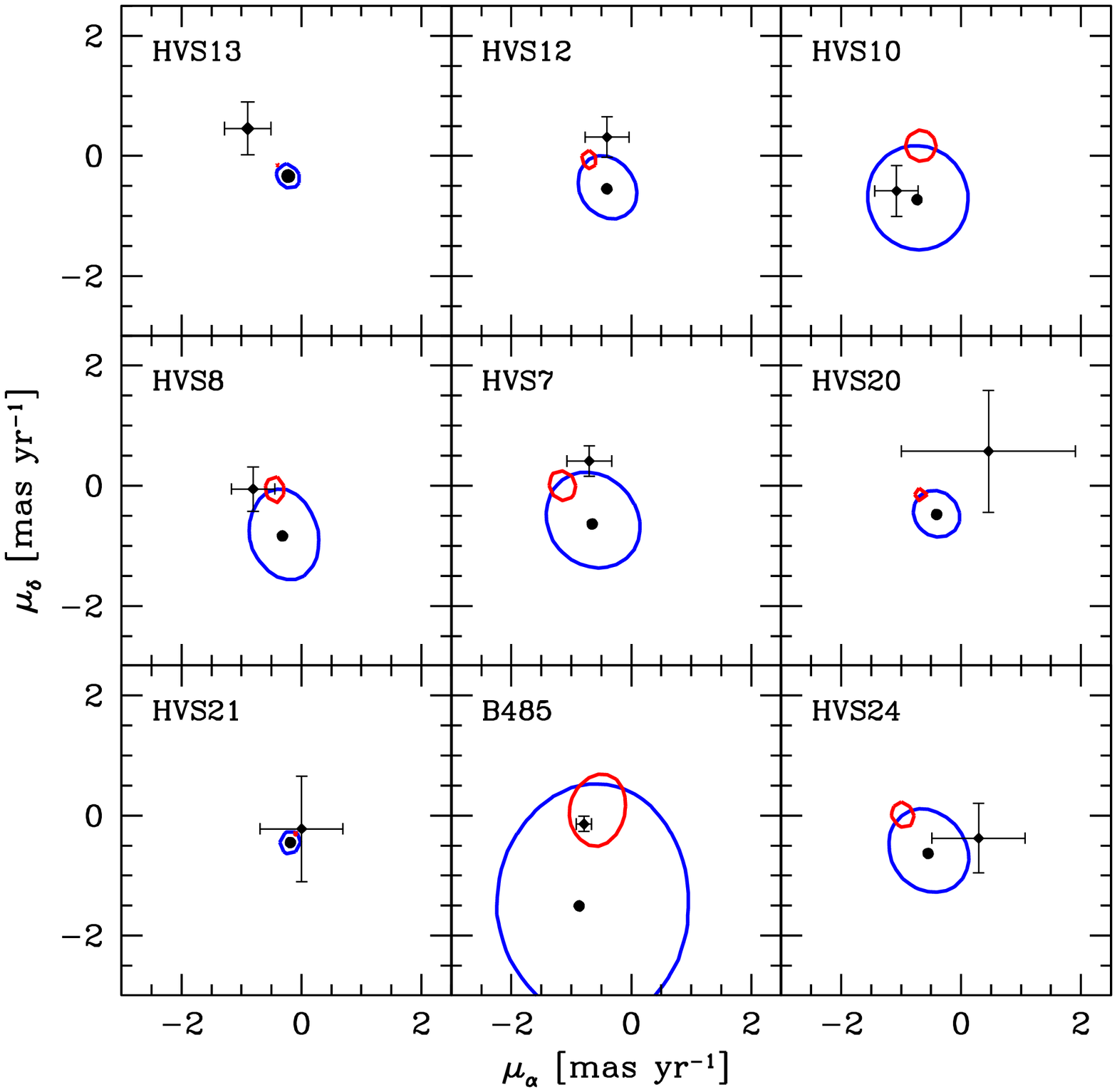} \\
    \noindent \includegraphics[width=3.5in]{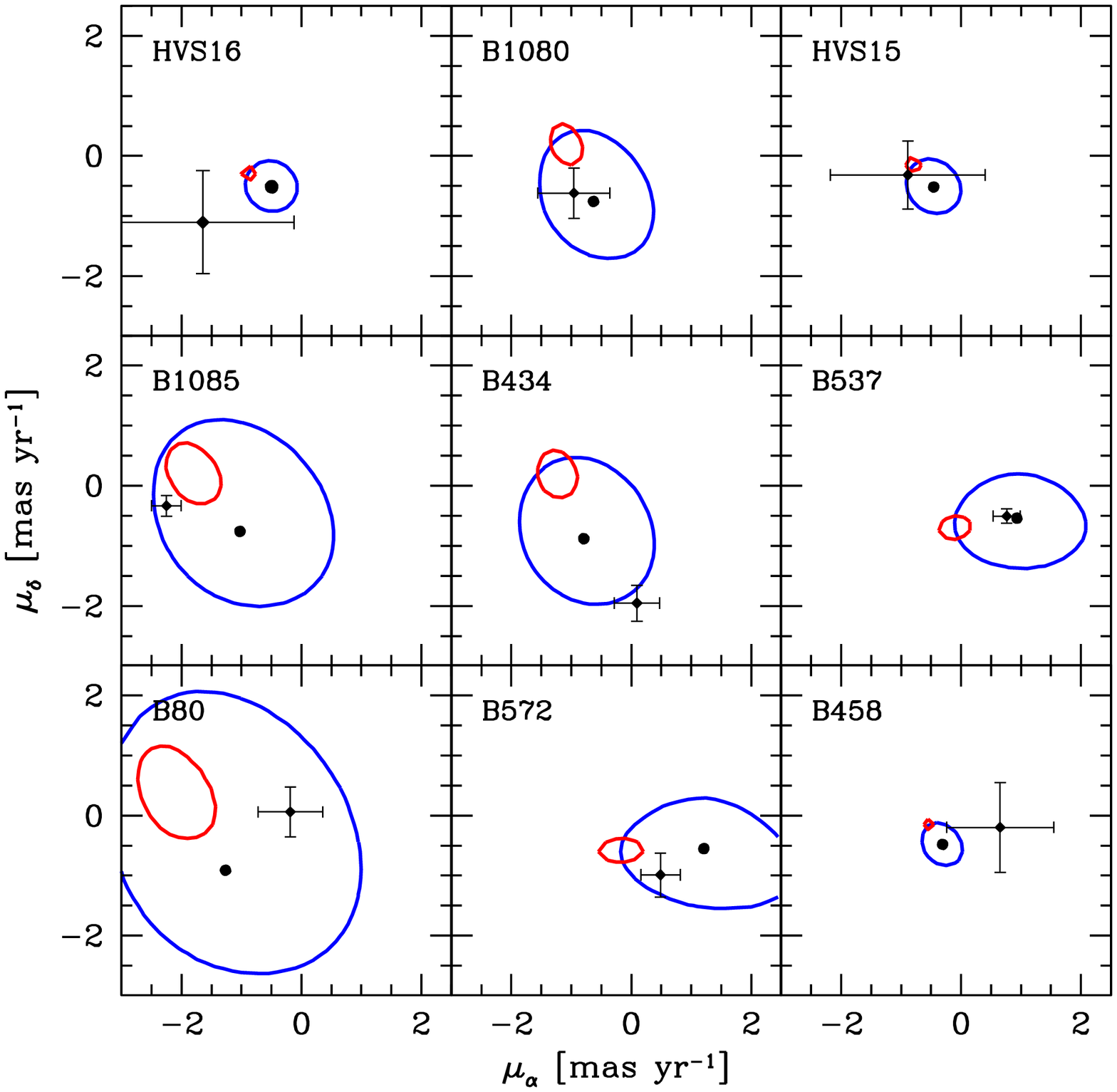}
	      \includegraphics[width=3.5in]{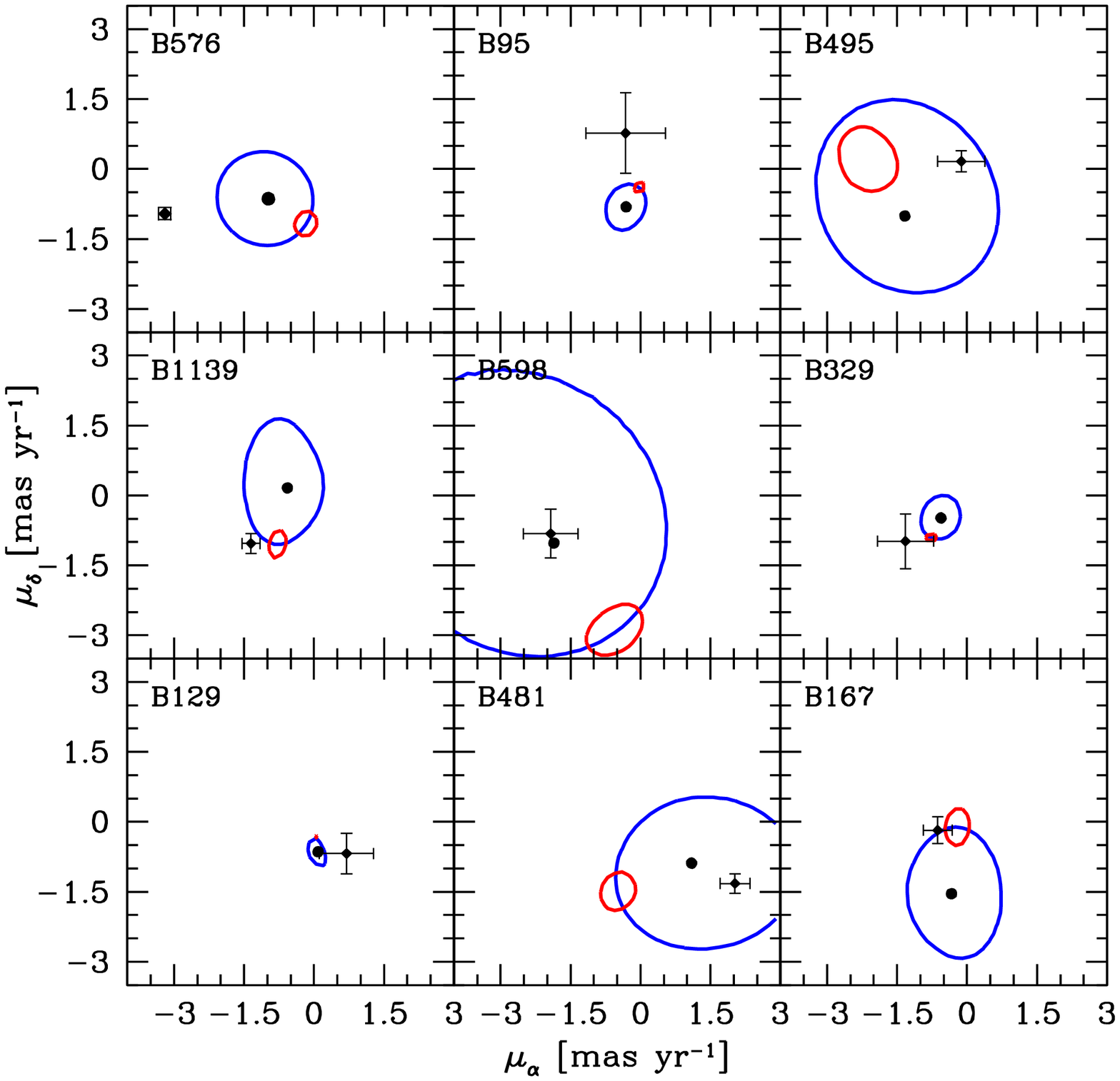} \\
    \centerline{ \includegraphics[width=3.5in]{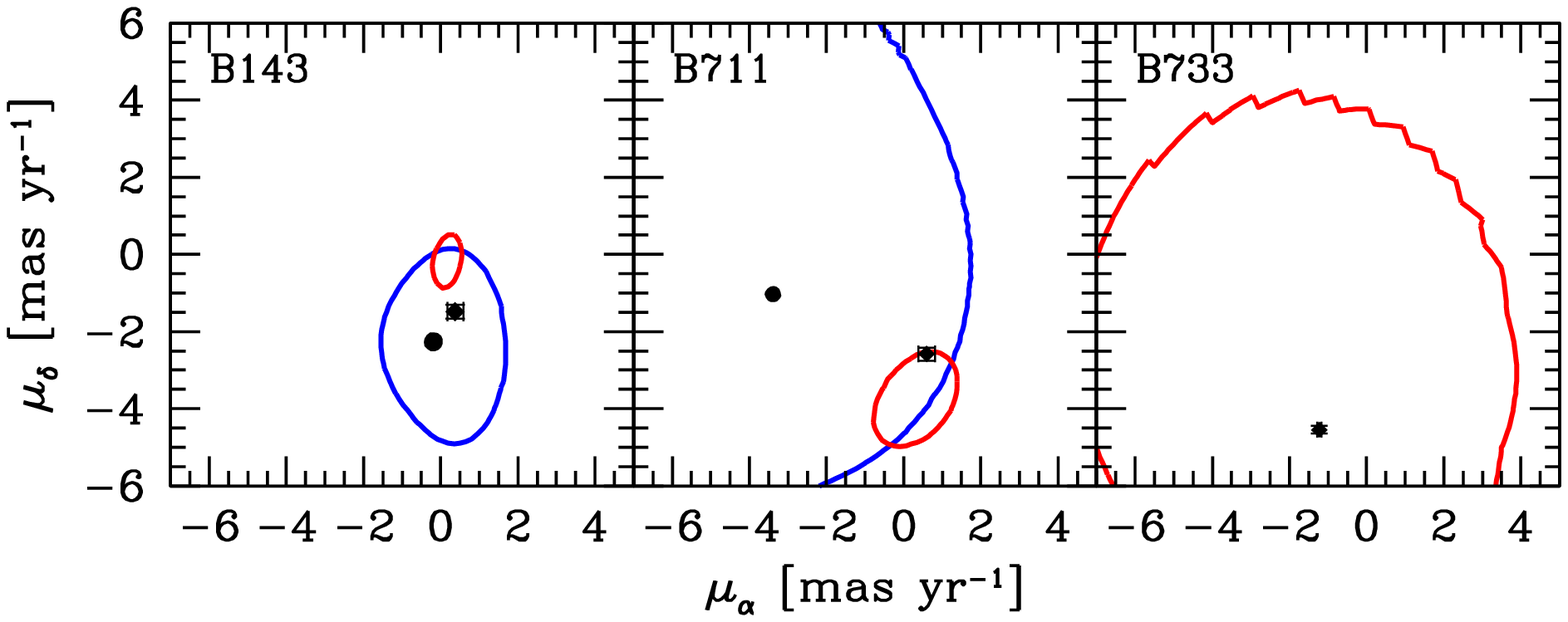} }
    \caption{ \label{fig:pm}
	Proper motion measurements and trajectories through the 
Milky Way.  Panels are ordered by decreasing $v_{rf}$, except for the final panel 
for the 3 nearest stars.  The blue ellipse is the locus of proper motions 
with trajectories that pass within 20 kpc of the Galactic Center, the edge of the 
Milky Way disk.  The small red ellipse is the locus of proper motions within 6 kpc 
of the point of minimum Galactic disk ejection velocity.  A Galactic Center 
trajectory is marked by the black dot.  }
 \end{figure*}

\subsection{Trajectory Calculation}

	Calculations are straightforward for the nearly radial trajectories of the
stars.  We start from the measured position and velocity of each star, and integrate
its trajectory backward in time through the Galactic potential.  We record where
each trajectory crosses the Galactic plane, and its distance from the Galactic
center.  We also record the ejection velocity required to launch the star from the
disk-crossing location, given the angle of the trajectory relative to Galactic
rotation.

	Figure \ref{fig:pm} plots the proper motion measurement for each star.  
Because we know position and radial velocity, a proper motion in Figure \ref{fig:pm}
corresponds to a unique trajectory through the Galaxy given a gravitational
potential model.  The blue ellipses in Figure \ref{fig:pm} are the loci of proper
motions with trajectories that cross the Galactic plane at $R=20$ kpc, the edge of
the Milky Way disk.  The red circle is the region of minimum Galactic disk ejection
velocity, the region where Galactic disk ejections can most easily escape the
gravitational pull of the Milky Way.  A Galactic Center trajectory is marked by the
large black dot.

	Measurement errors broaden the range of possible trajectories.  Thus we draw 
final conclusions from the density distribution of possible trajectories.  For each 
star, we perform $10^6$ Monte Carlo trajectory calculations assuming that 
measurement errors are normally distributed.  We generate correlated normal 
distributions for parallax, $\mu_{\alpha}$, $\mu_{\delta}$ using the {\it Gaia} 
correlation matrix and a Cholesky decomposition.   A $\pm$0.5 kpc uncertainty in 
solar position or a $\pm$10 \kms\ uncertainty in circular velocity yield 
negligible $\pm$0.5 kpc changes in Galactic plane crossing location and $\pm$0.2 
Myr changes in flight times.  Thus we set the solar position and circular velocity 
error to zero for simplicity.  Distance and proper motion are the two dominant 
sources of uncertainty.  We then measure the density of simulated trajectories as 
they cross the Galactic disk plane.

	We evaluate likelihood of origin based on the 0.3173 (1$\sigma$), 0.0455
(2$\sigma$), and 0.0027 (3$\sigma$) thresholds of the trajectory distribution at the
Galactic plane.  This approach is valid for testing the origin of unbound stars that
cross the Galactic plane only once, or bound stars with lifetimes less than their
orbital turn-around time; this approach is invalid for long-lived stars that cross
the plane more than once.  The measurements provide trajectory constraints for about
half of the sample, namely for the stars with $R<60$ kpc.

\subsection{Large Magellanic Cloud}

	Our northern hemisphere sample of stars is poorly suited to test the 
Large Magellanic Cloud origin hypothesis, which predicts a monopole of unbound stars 
in the southern sky \citep{boubert16, boubert17}.  Nearly all of our stars must pass 
{\it through} the disk of the Milky Way to reach the Large Magellanic Cloud in the 
south.  A better test is provided by the southern hemisphere star HE~0437$-$5439 
\citep{edelmann05}: its trajectory points from the Large Magellanic Cloud 
\citep{erkal18} and its unbound velocity possibly requires dynamical interaction 
with an intermediate mass black hole \citep{gualandris07}. 

\subsection{Ejection Velocity}

	The ejection velocity required to explain the present position and motion of 
our stars provides another constraint on their origin.  In the absence of a massive 
black hole, the speed limit for ejection from a stellar binary is set by the finite 
sizes of the stars \citep{leonard91}.  The orbital velocity of an equal-mass pair of 
stars separated by their radii is equal to the escape velocity from the surface of 
the stars.  Because stars on the main sequence have a quasi-linear relation between 
mass and radius, most stars in the Milky Way share a common escape velocity from 
their surface of about 600 \kms.  To achieve higher binary orbital speeds, main 
sequence stars would have to orbit inside each other, which is impossible.  A 600 
\kms\ speed limit is optimistic; the speed can only be lower if mass transfer, tidal 
heating, or binary evolution are taken into account \citep[e.g.][]{fregeau04, 
renzo18}.

	Chaining together dynamical and supernova ejections can theoretically 
yield a higher velocity \citep{pflamm10}, but the observable rate of such events is 
reduced by the joint probability of dynamically ejecting a binary and then 
disrupting it through a supernova explosion at maximum velocity in the same 
direction.  We estimate that the Galactic center ejection rate is orders of 
magnitude larger at $>$600 \kms\ speeds \citep{brown09a}.

	The ejection velocities required for the fastest stars in the HVS Survey
exceed 600 \kms.  Thus the ejection velocities demand a Galactic center origin.  
For stars near Galactic escape velocity, however, there is finite region of the
Milky Way disk where the stars can be ejected at $<$600 \kms.  The region is set by
the Milky Way gravitational potential and the rotation of the Milky Way disk:  the
ejection velocity minimum is located in the outer disk, at the position where the
disk rotation vector points in the direction of ejection \citep{bromley09}.  
Convolved with the power-law distribution of runaway ejection velocities
\citep{portegies00, perets12, renzo18}, the most probable disk runaway origin
location is this region of minimum ejection velocity.  We mark the minimum disk
ejection velocity region for each star in Figure \ref{fig:pm} with a red ellipse.

\section{Constraints on Origin}

\begin{figure}          % FIGURE:  RV Sample
    \includegraphics[width=3.25in]{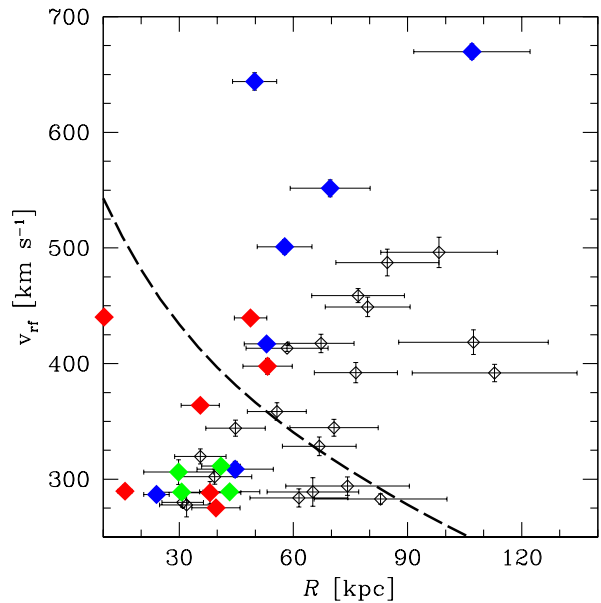}
    \caption{ \label{fig:vrad}
        Distribution of $v_{rf}$ versus Galactocentric radial distance $R$ for the 
39 velocity outliers with {\it Gaia} measurements.  Dashed line is Galactic escape 
velocity in our gravitational potential model \citep{kenyon14}.  Symbol color 
indicates probable origin:  Galactic center (blue), Galactic disk (red), Galactic 
Halo (green), and Ambiguous (empty). }
 \end{figure}

\begin{figure}          % FIGURE:  PVAL
    \includegraphics[width=3.4in]{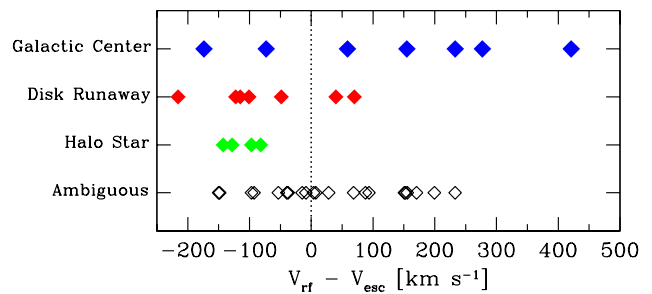}
    \caption{ \label{fig:pval}
	Probable origin, on the basis of trajectory and velocity, plotted relative
to Galactic escape velocity.  }
 \end{figure}

	We identify three classes of objects with distinct but overlapping velocity 
distributions in our sample:  1) Galactic center HVSs, 2) Galactic disk runaways, 
and 3) Galactic halo stars.  Figure \ref{fig:vrad} and Figure \ref{fig:pval} 
summarize the results.  We discuss the results in terms of $v_{rf}$, the 
heliocentric radial velocity transformed to the Galactic frame, because it is the 
largest component of velocity and the most accurately measured.  Figure 
\ref{fig:vrad} plots $v_{rf}$ versus Galactic radial distance $R$.  The dashed line 
is the Galactic escape velocity curve, and symbol color indicates the likely origin 
of each star.  Figure \ref{fig:pval} groups the origins together and plots them 
relative to Galactic escape velocity, $v_{rf} - v_{esc}$.  A total of 18 objects 
have robust constraints.

\subsection{Galactic Center Hypervelocity Stars} 

	A Galactic center origin is statistically preferred for all of the fastest
stars with $v_{rf}>+500$ km/s (HVS1, HVS4, HVS5, and HVS6).  The trajectories
currently provide 2$\sigma$ constraints.  The velocity itself provides an additional
physical constraint for the unbound stars:  the minimum Galactic disk ejection
velocity is comparable to the escape velocity from the surface of the stars, a
severe challenge to disk ejection.  We identify seven probable Galactic center HVS
ejections.

\begin{figure}          % FIGURE:  XY Bound HVS
    \includegraphics[width=3.5in]{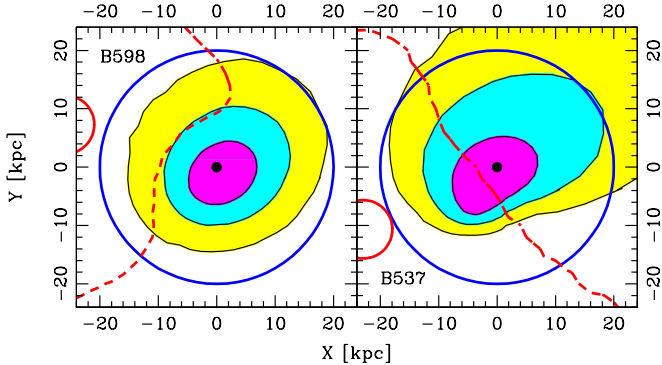}
    \caption{ \label{fig:xyh}
	Density of simulated trajectories (magenta region = 1$\sigma$, cyan region = 
2$\sigma$, yellow region = 3$\sigma$) where bound HVS candidates B537 and B598 cross 
the Galactic plane, in Cartesian coordinates.  The Galactic Center (black dot) is 
the origin most consistent with the measurements.  Trajectories from the region of 
minimum Galactic disk ejection velocity (solid red lines) are excluded at 3$\sigma$ 
confidence, however 500 \kms\ Galactic disk ejections (dashed red lines) are allowed.
	}
 \end{figure}

	Two bound stars, B537 and B598, have trajectories that point directly from 
the Galactic center, and reject the Galactic disk origin hypothesis at $>$3$\sigma$ 
significance.  To better understand these objects, Figure \ref{fig:xyh} plots the 
probability contours where these two objects cross the Galactic plane in Cartesian 
coordinates.  The solid red lines in Figure \ref{fig:xyh} mark the regions of 
minimum Galactic disk ejection velocity, excluded at 3$\sigma$ confidence.  The 
dashed red lines in Figure \ref{fig:xyh} mark the locus of trajectories with 500 
\kms\ Galactic disk ejection velocities.  Thus it is possible that B537 and B598 are 
high speed Galactic disk ejections.  A factor of 2 improvement in proper motion 
error would exclude this possibility for B598.  It is also possible that B537 and 
B598 are halo stars on very radial orbits.  High resolution spectroscopy can 
determine whether these are metal-poor halo stars or main sequence B stars.

	Extrapolating these results to the unconstrained half of the sample, we 
expect that about half of the unbound stars in the HVS Survey are Galactic center 
ejections.  \citet{brown14} thus overestimate the number of Galactic center 
ejections by a factor of two.  The implication is that, if HVSs are ejected 
continuously and isotropically, there are about 50 unbound 2.5-4 \msun\ HVSs over 
the entire sky to 100 kpc.  We compare this number with the theoretical predictions 
of \citet{zhang13}.  The models that best match the observed distribution of S-stars 
in the Galactic center and unbound stars in the Galactic halo predict about 10 to 30 
unbound 3-4 \msun\ HVSs over the entire sky.  Our revised number of unbound 
HVSs from the Galactic center is thus in better agreement with theoretical ejection 
calculations for the MBH ejection scenario.

\subsection{Disk Runaways} 

	We estimate the likelihood of Galactic disk origin by averaging the
trajectory density over a 5 kpc radius region centered on the position of minimum
Galactic disk ejection velocity.  This metric can be pessimistic for the nearest
bound objects, but it is well-matched to the average distance and proper motion
errors of the sample.
	We identify seven disk runaways with trajectories inconsistent with 
the Galactic center hypothesis at $\ge$3$\sigma$ significance, and statistically 
consistent with a Galactic disk ejection.

	Disk runaways and Galactic center HVSs have comparable frequency at speeds
around the Galactic escape velocity (Figure \ref{fig:pval}).  The fastest disk
runaways have $v_{rf}-v_{esc}\simeq+100$ \kms.  However, most disk runaways are 
bound.

	Interestingly, the two unbound disk runaways are spectroscopically 
unusual.  HVS7 and HVS17 are both chemically peculiar B stars \citep{przybilla08b, 
brown13b}.  HVS7 is 10-100 times under-abundant in He and in CNO relative to the 
Sun, and 100-10,000 time over-abundant in iron group and rare-earth elements 
\citep{przybilla08b}.  It is unclear whether abundance patterns are linked to a 
supernova binary disruption origin, however.  The abundance patterns of the unbound 
runaway B stars HVS7, HVS17, and HD~271791 \citep{przybilla08c} differ 
significantly.  \citet{mcevoy17} find no correlation in a more detailed 
abundance analysis of 38 runaway B stars.

	The frequency of unbound runaways is linked to their ejection rate.  
Theoretical ejection models predict a power-law distribution of ejection velocities
\citep{portegies00, perets12, renzo18}.  In these models, $>$99\% of runaways are
launched from the disk with $<$200 \kms\ speeds.  Simulated distributions of
runaways in the Milky Way therefore predict a flattened spatial distribution of
runaways with a scale height comparable to the thick disk \citep{bromley09}.  We
expect that magnitude-limited surveys of less luminous types of stars than B stars
will find many more runaways near the disk.

\subsection{Halo stars} 

	Halo stars first appear at $v_{rf}<+300$ \kms\ in our sample.  There are 
four bound objects with trajectories that significantly reject both Galactic center 
and Galactic disk origins, and which cross the disk in the direction opposite 
Galactic rotation.  These stars are likely halo stars, presumably hot blue 
horizontal branch stars with temperatures and surface gravities similar to main 
sequence B stars.  The number of halo stars is consistent with the number of 
$-300<v_{rf}<-275$ \kms\ negative velocity outliers in the HVS Survey.  At even 
lower $v_{rf}<250$ \kms\ velocities, halo stars completely dominate the HVS Survey 
\citep{brown14}.

	Indeed, the {\it Gaia} measurements show that 94\% of previously 
claimed unbound stars are likely bound halo stars \citep{boubert18}.  Searches 
targeting high velocity stars in {\it Gaia} \citep{marchetti18, hattori18b} 
predominantly find low-mass, metal-poor stars moving in equal numbers towards and 
away from the Sun; in other words, halo stars. 

\section{Conclusions}

	{\it Gaia} proper motions enable distinction between true Galactic center 
HVSs and other high velocity stars.  We examine the probable origin for B-type stars 
from the HVS Survey.  Eighteen objects have robust constraints.  Halo stars dominate 
the sample at bound speeds, $v_{rf}-v_{esc}<-100$ \kms.  We identify seven disk 
runaways with trajectories that significantly reject the Galactic center hypothesis.  
The fastest disk runaways have $v_{rf}-v_{esc}\simeq+100$ \kms, but the majority are 
bound.  We identify seven probable Galactic center HVS ejections.  The Galactic 
center ejections dominate the sample at speeds significantly above Galactic escape 
velocity.

	A clean sample of HVSs is important for constraining the Galactic center 
ejection mechanism.  With a sample of 50 HVSs, for example, discrimination between a 
single and a binary MBH ejection scenario might be possible \citep{sesana07b}.  
Different MBH ejection mechanisms predict different spatial distributions of HVSs on 
the sky \citep{levin06, abadi09, zhang13, boubert16, kenyon18}.  These distributions 
can be tested with larger samples of HVSs.

	A well-defined sample of HVSs will also enable measurement of the ejection
rate of stars encountering the MBH.  Tidal disruption events may be closely related
to the encounters that produce HVSs \citep{bromley12}.

	{\it Gaia} end-of-mission proper motion errors should improve by a factor of
3.  There are a dozen objects in our sample with 1-2$\sigma$ trajectory constraints
that will have $>$3$\sigma$ constraints in only a few years.  For the faintest
stars, however, {\it Gaia} cannot compete with pointed observations.  Next
generation missions like the proposed {\it Theia} mission \citep{theia17} are needed
to measure HVS proper motions with uncertainties significantly below 0.1 \mas.

	With uncertainties below 0.1 \mas, HVSs become important tools for measuring
the Milky Way mass distribution \citep{gnedin05, yu07}.  HVSs integrate the
gravitational potential from the very center of the Milky Way to its outermost
regions.  Theorists have proposed measuring the angular momentum of one very
nearby HVS \citep{hattori18a} or using the phase space distribution of hundreds of
HVSs \citep{perets09b, rossi17, contigiani18} to constrain the Milky Way mass
distribution.  Thus HVSs can complement studies possible now, that use halo star
streams \citep{bonaca18}, globular clusters \citep{posti18, watkins18}, or dwarf
galaxies \citep{fritz18} to constrain the Milky Way dark matter halo.  Any deviation
of a HVS's trajectory from the Galactic center measures the Milky Way's
non-spherical mass distribution, independent of any other technique.

\facility{Gaia}

	~

\acknowledgements
	This work has made use of data from the European Space Agency mission {\it 
Gaia}, processed by the {\it Gaia} Data Processing and Analysis Consortium (DPAC).  
Funding for the DPAC has been provided by national institutions, in particular the 
institutions participating in the {\it Gaia} Multilateral Agreement.  This research 
has made use of NASA's Astrophysics Data System.  This research was supported in 
part by the Smithsonian Institution.

% TABLE 2
\begin{deluxetable*}{lcccccc}
\tabletypesize{\scriptsize}
\tablecaption{Constraints on Origin\label{tab:pval}}
\tablewidth{0pt}
\tablecolumns{7}
\tablehead{
  \colhead{ID} & \colhead{$v_{rf}$} & \colhead{{\it R}} & \colhead{$v_{rf}-v_{esc}$} & 
  \colhead{$p_{\rm GC~HVS}$} & \colhead{$p_{\rm Disk~Runaway}$} \\
  \colhead{} & \colhead{\kms } & \colhead{kpc} & \colhead{\kms } &
  \colhead{ } & \colhead{ }
}
        \startdata
HVS1	& $669.8\pm 6.6$ & $106.9\pm15.3$ & $ 421$ & 0.41959 & 0.13256 \\
HVS5	& $644.0\pm 7.5$ & $ 49.8\pm 5.8$ & $ 277$ & 0.02566 & 0.00234 \\
HVS4	& $551.7\pm 7.3$ & $ 69.7\pm10.5$ & $ 233$ & 0.80644 & 0.02091 \\
HVS6	& $501.1\pm 6.3$ & $ 57.7\pm 7.2$ & $ 155$ & 0.60588 & 0.08691 \\
HVS19	& $496.2\pm13.1$ & $ 98.3\pm15.3$ & $ 233$ & 0.84863 & 0.72447 \\
HVS22	& $487.4\pm11.5$ & $ 84.7\pm13.5$ & $ 199$ & 0.06899 & 0.06631 \\
HVS9	& $458.8\pm 6.1$ & $ 77.0\pm12.2$ & $ 156$ & 1.00000 & 0.59219 \\
HVS18	& $449.0\pm 8.5$ & $ 79.5\pm11.1$ & $ 151$ & 0.79264 & 0.93030 \\
B733	& $440.3\pm 2.3$ & $ 10.2\pm 1.2$ & $-101$ & 0.00000 & 0.00016 \\
HVS17	& $439.5\pm 4.6$ & $ 48.7\pm 4.3$ & $  70$ & 0.00036 & 0.07498 \\
HVS13	& $418.5\pm10.8$ & $107.3\pm19.6$ & $ 170$ & 0.12816 & 0.84341 \\
HVS12	& $417.4\pm 8.1$ & $ 67.2\pm 8.7$ & $  94$ & 0.06419 & 0.36458 \\
HVS10	& $417.0\pm 4.6$ & $ 53.0\pm 5.9$ & $  59$ & 0.92961 & 0.10883 \\
HVS8	& $413.3\pm 2.6$ & $ 58.3\pm10.8$ & $  69$ & 0.43309 & 0.52037 \\
HVS7	& $397.7\pm 6.8$ & $ 53.2\pm 6.5$ & $  40$ & 0.00100 & 0.11113 \\
HVS20	& $392.1\pm 8.7$ & $ 76.4\pm10.9$ & $  88$ & 0.22497 & 0.16428 \\
HVS21	& $391.9\pm 7.5$ & $112.9\pm21.7$ & $ 153$ & 0.87222 & 0.54908 \\
B485	& $363.9\pm 4.8$ & $ 35.5\pm 5.0$ & $ -48$ & 0.00000 & 0.28713 \\
HVS24	& $358.6\pm 7.6$ & $ 55.7\pm 7.7$ & $   7$ & 0.53504 & 0.16045 \\
HVS16	& $344.6\pm 7.3$ & $ 70.7\pm11.6$ & $  28$ & 0.98818 & 0.65225 \\
B1080	& $344.1\pm 6.9$ & $ 44.8\pm 7.8$ & $ -37$ & 1.00000 & 0.21767 \\
HVS15	& $328.5\pm 8.1$ & $ 66.8\pm 9.7$ & $   4$ & 1.00000 & 0.73678 \\
B1085	& $319.6\pm 6.5$ & $ 35.5\pm 6.8$ & $ -93$ & 0.22286 & 0.01703 \\
B434	& $311.4\pm 2.9$ & $ 41.0\pm 5.1$ & $ -82$ & 0.00155 & 0.00000 \\
B537	& $308.6\pm 6.9$ & $ 44.7\pm10.0$ & $ -73$ & 0.72596 & 0.00000 \\
B080	& $306.2\pm10.7$ & $ 29.8\pm 9.1$ & $-128$ & 0.00036 & 0.02590 \\
B572	& $302.1\pm 6.5$ & $ 39.3\pm 9.7$ & $ -97$ & 0.16757 & 0.00489 \\
B458	& $294.1\pm 7.9$ & $ 74.2\pm16.3$ & $ -15$ & 0.46156 & 0.13478 \\
B711	& $289.6\pm 5.4$ & $ 15.7\pm 1.9$ & $-216$ & 0.00000 & 0.04737 \\
B576	& $289.1\pm 2.4$ & $ 43.2\pm 7.9$ & $ -97$ & 0.02239 & 0.00000 \\
B095	& $289.1\pm12.3$ & $ 65.1\pm12.0$ & $ -40$ & 0.41893 & 0.97142 \\
B495	& $288.9\pm 4.6$ & $ 30.6\pm 5.6$ & $-142$ & 0.00000 & 0.00243 \\
B1139	& $288.3\pm 9.7$ & $ 38.1\pm 8.1$ & $-115$ & 0.00000 & 0.02608 \\
B598	& $286.8\pm 5.7$ & $ 24.0\pm 3.3$ & $-174$ & 0.93866 & 0.00000 \\
B329	& $283.9\pm 7.9$ & $ 61.4\pm12.8$ & $ -53$ & 1.00000 & 0.41425 \\
B129	& $282.8\pm 4.5$ & $ 82.9\pm17.5$ & $  -9$ & 0.21733 & 0.02988 \\
B143	& $279.9\pm 4.4$ & $ 30.9\pm 5.5$ & $-150$ & 0.02091 & 0.00000 \\
B481	& $277.8\pm10.4$ & $ 31.9\pm 7.1$ & $-148$ & 0.36545 & 0.00000 \\
B167	& $275.2\pm 4.3$ & $ 39.7\pm 6.3$ & $-123$ & 0.00000 & 0.47101 \\
	\enddata
\end{deluxetable*}

\end{document}